\documentclass{article}
\pdfoutput=1
\usepackage[nonatbib,preprint]{neurips_2019}

\usepackage[utf8]{inputenc} % allow utf-8 input
\usepackage[T1]{fontenc}    % use 8-bit T1 fonts
\usepackage{hyperref}       % hyperlinks
\usepackage{url}            % simple URL typesetting
\usepackage{booktabs}       % professional-quality tables
\usepackage{amsfonts}       % blackboard math symbols
\usepackage{nicefrac}       % compact symbols for 1/2, etc.
\usepackage{microtype}
\usepackage{graphicx}% microtypography

\usepackage{multirow}
\usepackage[table,xcdraw]{xcolor}

\title{Quantifying the Carbon Emissions \\ of Machine Learning}

\author{%
  Alexandre Lacoste\thanks{equal contribution}\\
  Element AI\\
  \texttt{allac@elementai.com} \\
  \And
  Alexandra Luccioni\footnotemark[1]\\
  Mila, Université de Montréal\\
  \texttt{luccionis@mila.quebec} \\
  \AND
  Victor Schmidt\footnotemark[1]\\
  Mila, Université de Montréal\\
  \texttt{schmidtv@mila.quebec} \\
  \And
  Thomas Dandres \\
 Polytechnique Montréal, CIRAIG\\
  \texttt{thomas.dandres@polymtl.ca} \\
}

\begin{document}

\maketitle

\begin{abstract}
From an environmental standpoint, there are a few crucial aspects of training a neural network that have a major impact on the quantity of carbon that it emits. These factors include: the location of the server used for training and the energy grid that it uses, the length of the training procedure, and even the make and model of hardware on which the training takes place. In order to approximate these emissions, we present our \emph{\href{https://mlco2.github.io/impact/}{Machine Learning Emissions Calculator}}, a tool for our community to better understand the environmental impact of training ML models. We accompany this tool with an explanation of the factors cited above, as well as concrete actions that individual practitioners and organizations can take to mitigate their carbon emissions.
\end{abstract}

\section{Introduction}

While a decade ago, only a few ML pioneers were training neural networks on GPUs (Graphical Processing Units), in recent years powerful GPUs have become increasingly accessible and used by ML practitioners worldwide. Furthermore, new models often need to beat existing challenges, which entails training on more GPUs, with larger datasets, for a longer time. This expansion brings with it ever-growing costs in terms of the energy needed to fuel it. This trend has been the subject of recent studies aiming to evaluate the climate impact of AI, which have predominantly put the focus on the environmental cost of training large-scale models connected to grids powered by fossil fuels~\cite{strubell, greenai}. While these models are not necessarily representative of common practice, we believe that it is important to continue this conversation further and work towards defining the tools and steps that we need to assess the carbon emissions generated by the models we train, as well as to propose ways to reduce those emissions.

In this work, we present our Machine Learning Emissions Calculator (\url{https://mlco2.github.io/impact/}),  a tool for our community to estimate the amount of carbon emissions produced by training ML models. We accompany this tool with a presentation of key concepts and an explanation of the factors impacting emissions. % We then illustrate the way in which total network training emissions can vary widely %in a case study of a few standard ML model architectures depending on deployment choices.
Finally, we end our article with some recommendations of best practices for the overall ML research community, as well as for individual researchers.

\section{Quantifying Carbon Emissions in Neural Network Training}\label{quantif}

In order to quantify carbon emissions, we use \textit{CO\textsubscript{2}-equivalents} (CO\textsubscript{2}eq), which is a standardized measure used to express the global-warming potential of various greenhouse gases as a single number, i.e. as the amount of CO\textsubscript{2} which would have the equivalent global warming impact~\cite{ipcc2006}. We will use this single metric to compare the factors and choices that impact overall amount of emissions produced by training an ML model in the sections below.

\subsection{Type of Energy Used}\label{energy}

\begin{figure}[h!]
    \centering
    \includegraphics[bb=0 0 672 275, width=0.7\textwidth]{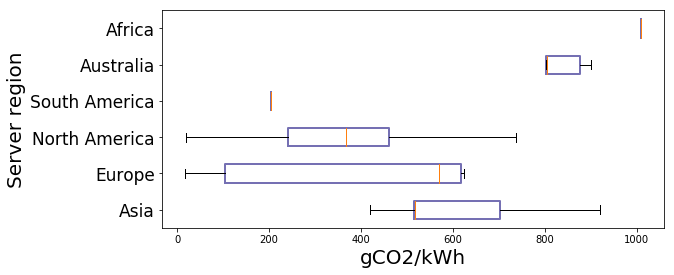}
    \caption{Variation of the Average Carbon Intensity of Servers Worldwide, by Region. (Vertical bars represent regions with a single available data point.) }
    \label{energy_plot}
\end{figure}

Practically speaking, it is hard to estimate exactly the amount of CO\textsubscript{2}eq emitted by a cloud server in a given location because the information regarding the energy grid that it is connected to is rarely publicly available. However, if we assume that all servers are connected to local grids at their physical location, we are able to make an estimation of the amount of CO\textsubscript{2}eq that they emit using public data sources~\cite{hongkong, emissionsfactors} . Therefore, in order to create our emissions calculator, we gathered data regarding CO\textsubscript{2}eq emissions of different grid locations and cross-referenced them with known GPU server locations from the three major cloud providers: Google Cloud Platform, Microsoft Azure and Amazon Web Services \footnote{The data can be found at~\url{https://github.com/mlco2/impact/tree/master/data}}. Our aim in doing this is to illustrate the degree of variability that exists depending on the location of a given server. For instance, in Figure~\ref{energy_plot}, we show the distribution and variation in carbon emissions depending on geographical region. It can be noted that a large amount of variation can be found within a single region; for instance, servers located in North America can emit anywhere between 20g CO\textsubscript{2}eq/kWh in Quebec, Canada to 736.6g CO\textsubscript{2}eq/kWh in Iowa, USA~\cite{emissionsfactors}.

\subsection{Computing Infrastructure and Training Time}
Another, more subtle, factor in carbon emitted by a neural network is the computing infrastructure used and training time of the model. In terms of performance the number of floating point operations per second (FLOPS) of GPUs has been steadily increasing in recent years, from 100 Giga FLOPS per second in 2004 to up to 15 Tera FLOPS per second in recent hardware~\cite{gpu}. However, with neural network architectures becoming deeper and more complex, recent state-of-the-art models are often trained on multiples GPUs for several weeks (or months) to beat benchmark performance, requiring more and more energy.

Finally, when it comes to defining a training procedure for ML architectures, there are several elements to consider: for starters, whether it is necessary to train a model from scratch or whether fine-tuning is adequate for the task at hand. Notably, recent research has shown that using pre-trained models with task-specific fine-tuning performs as well as training from scratch, while being more robust, for tasks in image recognition~\cite{cnn,foodrecognition} and NLP~\cite{ulmfit}. Furthermore, when it comes to hyperparameter search, it has been proven both empirically and theoretically that random hyperparameter search is more efficient than grid search for hyperparameter optimization~\cite{random}, and there is much research being done on ways to improve the efficiency of hyperparameter optimization~\cite{bohb,hyperparameter}, which makes it possible to continue choosing the right hyperparameters for new models without incurring superfluous computing and energy costs.

\section{ML Emissions Calculator and Actionable Items}

It is difficult to provide clear-cut guidelines for ML researchers to follow in order to reduce the carbon emissions, or specific benchmarks for the training time that a given model or task warrants. Nonetheless, we think that there are certain best practices and actionable items that can be adopted by our community to reduce environmental impact of the ML domain. We present some of these, along with our ML emissions calculator, in the current section.

 \paragraph{Quantify Your Emissions} Being informed regarding the factors that impact the quantity of carbon emissions produced by ML research is the first step to making positive changes. It is for this reason that we created our \emph{ML Emissions Calculator}. This tool, currently in its alpha version, takes as input the details regarding the training of an ML model: the geographical zone of the server, the type of GPU, and the training time, and gives as output the approximate amount of CO\textsubscript{2}eq produced. We collected publicly available data for the 4 main variables of this computation: (i) the energy consumption of hardware (see "Choose More Efficient Hardware" below), (ii) the location of providers' regions of compute -- which we assumed to be connected to their local grid, (iii) the region's CO\textsubscript{2}eq emissions per kWh and (iv) potential offsets bought by the provider.

 We intend to adopt an open and transparent approach: the data we used is publicly available, debatable and editable through Github issues and pull requests. We are therefore open to updating data as more information becomes available. Since this paper's core goal is to raise awareness around the carbon emissions of ML, we have also included two educational sections in the website: one about learning the main notions and concepts related to this domain (e.g. RECs, carbon neutrality, etc.), the other about actionable items an individual or an organization can leverage to mitigate their carbon impact.

\paragraph{Choose Cloud Providers Wisely} In recent years, many cloud providers have defined ambitious sustainability goals and are offsetting their emissions through Renewable Energy Certificates (RECs) in an effort to become carbon neutral, a term used to indicate a net zero carbon footprint of an organization. Each REC bought attests that 1 MWh of renewable energy has been added to the energy grid and can be used to offset an equivalent amount of non-renewable energy. For instance, Google Cloud Platform is certified carbon neutral and funds solar and wind farms directly on local grids through RECs~\cite{google}. Microsoft Azure is also certified carbon neutral and 44\% of its electricity consumption directly comes from renewable energy, according to a 2016 estimate~\cite{microsoft}. Finally, to the best of our knowledge, while not yet 100\% carbon neutral on an organizational level, Amazon Web Services is also funding renewable energy projects and some of their data centers are powered by renewable energy~\cite{aws}.

Another major energy consumption factor of server installations is the power usage effectiveness (PUE) of the centers where the GPUs are hosted, which represents the percentage of energy consumption that is used for cooling, power conversion, and other auxiliary tasks, and can vary immensely. For example, Google Cloud Services has an average PUE of 1.1, meaning that only 11\% of their total energy usage is not used for the servers themselves, a ratio that they have been steadily reducing using Reinforcement Learning~\cite{gao, googledata}.
Finally, if you rely on a local private compute infrastructure, it is also possible to engage with administrators about quantifying and offsetting the emissions produced, as well as improving the efficiency of your grid -- this may help bring your organization toward carbon neutrality and have a significant impact at scale.

\paragraph{Select Data Center Location} While many cloud providers are carbon neutral, some of their data centers may still be carbon intensive due to the local grid that they are connected to, whereas others will be low carbon and powered solely by renewable energy sources. Hence, selecting the data center location where an algorithm will be trained has a large impact on its direct carbon emissions. This choice can be achieved by consciously selecting the server location before dispatching your jobs. As we illustrated in previous sections, this single choice can make the direct emissions of an algorithm vary by a factor of 40, from 20g CO\textsubscript{2}eq/kWh in a location that uses renewable energy sources to 820g CO\textsubscript{2}eq/kWh in a location that solely relies on fossil fuels~\cite{emissionsfactors}. For a model such as VGG~\cite{vgg} or BERT~\cite{bert}, which are trained on multiple GPUs for several weeks, this can correspond to avoiding emitting several hundreds of kilograms of CO\textsubscript{2}eq by training on a server powered by hydroelectricity instead of fossil fuels.

\paragraph{Reduce Wasted Ressources} Grid search is still often used in practice, in spite of its low efficiency both in terms of model performance and environmental impact. However, it has been shown that random search (and others) not only is a straightforward replacement but also has potential to significantly accelerate hyperparameter search~\cite{li2016hyperband, li2018massively, falkner2018bohb}, consequently reducing carbon emissions. Also, while failed experiments are a common part of ML research and are sometimes unavoidable, their number can often be reduced with careful design such as unit tests, integration tests, and extensive and early debugging. Uninformative experiments are also frequent (sometimes unknowingly) -- they can be caused by unstable learning algorithms requiring averaging results over many random seeds. Taking the time to carry out a literature review and to understand the potential sources of noise before launching large-scale hyperparameter searches increases the chance of obtaining reproducible and statistically significant results. Hence, reducing the need to extend the experiment cycles.

\paragraph{Choose More Efficient Hardware} The choice of computing hardware can also have a major impact on ML emissions. To perform a comparison between different devices, their compute efficiency can be estimated in FLOPS/W. This estimation is based on their theoretical peak performance with respect to their Thermal Design Power (TDP)\footnote{Empirical measurement of GFLOPS/W on various ML architecture would provide more accurate numbers but we are only interested in approximate values to compare classes of devices.}. Using this approach, it can be found that CPUs can be 10 times less efficient than GPUs while TPU 3 can be 4 to 8 times more efficient than GPUs \cite{TPU} (refer to Table~\ref{tab:hardware-efficiency} for details). Interestingly, in contexts where low power consumption and efficiency are important, e.g., for embedded applications, GPUs such as the Jetson AGX Xavier can be 10 to 20 times more efficient than traditional GPUs.

\section{Discussion}

The factors that we discussed in the current work give ML practitioners a certain amount of control over the environmental impact produced by the training of their models. We are aware that these choices are not always possible to make in practice -- for instance, the choice of server location can be limited due to privacy considerations in the case of applications in the medical or financial domain, and large amounts of data may be needed to produce most robust models. However, we find that our emissions calculator is a good starting point to estimate the impact that small choices in model training can have on direct carbon emissions resulting from ML research.

Despite our best efforts, our calculator remains simply an approximation of the true emissions produced by ML training for several reasons: to start with, there is the issue of global load balancing, i.e. if a majority of practitioners choose to run their models in a low-carbon location, the servers will get saturated and other servers will still need be used.
In that perspective, the global gain will not be a 40-fold reduction of emissions, but much smaller. Furthermore, there is a lack of transparency with regards to the true quantity of emissions produced by organizations, so while we use the current best publicly-available sources, there is still a large margin of error with regards to the exact quantity of energy consumed and carbon produced -- we remain open to additional data sources and numbers. Finally, while in the current version of our tool, we focus on quantifying the emissions of training ML models, there is still the issue of deploying them, since the inference process is also energy-expensive, especially if done continuously and on a large scale. This is something that should be taken into account by ML practitioners in their products that are deployed in real-world settings, for instance by using energy-efficient architectures~\cite{neuralpower} and computing infrastructure.

There are also more far-reaching discussions to be had regarding the environmental value of scientific knowledge in general and of ML research in particular. On one hand, there is valuable research to be done in ML especially with regards to tackling climate change~\cite{tackling, visualizing}, whereas on the other hand, the emissions of the field of ML are growing quickly~\cite{strubell}. We do not propose the solution to this problem, but we believe that there are steps to be taken, for instance by using efficiency as an evaluation criterion (as proposed by~\cite{greenai}) or by taking concrete steps to reduce emissions (as proposed by the current paper). We hope that our work, along with others, will open the door for these conversations and debates to take place, to quantify the environmental impact of our field, and for positive changes that can be made to reduce it.

\newpage
\bibliographystyle{unsrt}
\bibliography{references}

\newpage

\section*{Appendix A: Energy Grid Data Used for the ML Emissions Calculator}\label{appendixA}

For clarity purposes, the data presented in this appendix contains fewer columns than what can be found in our public database: https://github.com/mlco2/impact/tree/master/data. For instance we did not include the sources in these tables. For future reference, the data's commit hash at the time of publication is e692e28.

\subsection*{Google Cloud Platform}

\begin{table}[h!]
\small
\centering

\begin{tabular}{lllc}
\rowcolor[HTML]{EFEFEF}
{\color[HTML]{333333} \textbf{Region}} & {\color[HTML]{333333} \textbf{Country}} & {\color[HTML]{333333} \textbf{City}} & {\color[HTML]{333333} \textbf{Estimated gCO2e/kWh)}} \\ \hline
 & & & \\
asia-east1                                  & Taiwan                                  & Changhua County                      & 557                                            \\
asia-east2                                  & China                                   & Hong Kong                            & 702                                            \\
asia-northeast1                             & Japan                                   & Tokyo                                & 516                                            \\
asia-northeast2                             & Japan                                   & Osaka                                & 516                                            \\
asia-south1                                 & India                                   & Mumbai                               & 920                                            \\
asia-southeast1                             & Singapore                               & Jurong West                          & 419                                            \\
australia-southeast1                        & Australia                               & Sydney                               & 802                                            \\
europe-north1                               & Finland                                 & Hamina                               & 211                                            \\
europe-west1                                & Belgium                                 & St. Ghislain                         & 267                                            \\
europe-west2                                & United Kingdom                          & London                               & 623                                            \\
europe-west3                                & Germany                                 & Frankfurt                            & 615                                            \\
europe-west4                                & Netherlands                             & Eemshaven                            & 569                                            \\
europe-west6                                & Switzerland                             & Zürich                               & 16                                             \\
northamerica-northeast1                     & Canada                                  & Montréal                             & 20                                             \\
southamerica-east1                          & Brazil                                  & São Paulo                            & 205                                            \\
us-central1                                 & USA                                     & Council Bluffs                       & 566.3                                          \\
us-east1                                    & USA                                     & Moncks Corner                        & 367.8                                          \\
us-east4                                    & USA                                     & Ashburn                              & 367.8                                          \\
us-west1                                    & USA                                     & The Dalles                           & 297.6                                          \\
us-west2                                    & USA                                     & Los Angeles                          & 240.6
\end{tabular}
\end{table}

\subsection*{Amazon Web Services}

\begin{table}[h!]
\centering
\small
\begin{tabular}{llll}
\rowcolor[HTML]{EFEFEF}
\textbf{Region} & \textbf{Country} & \textbf{City}     & \textbf{gCO2e/kWh} \\ \hline
 & & & \\
us-east-2       & USA              & Columbus          & 568.2              \\
us-east-1       & USA              & Ashburn           & 367.8              \\
us-west-1       & USA              & San Francisco     & 240.6              \\
us-west-2       & USA              & Portland          & 297.6              \\
ap-east-1       & China            & Hong Kong         & 702                \\
ap-south-1      & India            & Mumbai            & 920                \\
ap-northeast-3  & Japan            & Osaka             & 516                \\
ap-northeast-2  & South Korea      & Seoul             & 517                \\
ap-southeast-1  & Singapore        & Singapore         & 419                \\
ap-southeast-2  & Australia        & Sydney            & 802                \\
ap-northeast-1  & Japan            & Tokyo             & 516                \\
ca-central-1    & Canada           & Montreal          & 20                 \\
cn-north-1      & China            & Beijing           & 680                \\
cn-northwest-1  & China            & Zhongwei          & 680                \\
eu-central-1    & Germany          & Frankfurt am Main & 615                \\
eu-west-1       & Ireland          & Dublin            & 617                \\
eu-west-2       & United Kingdom   & London            & 623                \\
eu-west-3       & France           & Paris             & 105                \\
eu-north-1      & Sweden           & Stockholm         & 47                 \\
sa-east-1       & Brazil           & Sao Paulo         & 205                \\
us-gov-east-1   & USA              & Dublin            & 568.2              \\
us-gov-west-1   & USA              & Seattle           & 297.6
\end{tabular}
\end{table}

\newpage
\subsection*{Microsoft Azure}

\begin{table}[h!]
\small
\centering
\begin{tabular}{llll}
\rowcolor[HTML]{EFEFEF}
\textbf{Region}    & \textbf{Country} & \textbf{City} & \textbf{gCO2e/kWh} \\ \hline
 & & & \\
eastasia           & Hong Kong        & Wan Chai      & 702                \\
southeastasia      & Singapore        & Singapore     & 419                \\
centralus          & USA              & Des Moines    & 736.6              \\
eastus             & USA              & Blue Ridge    & 367.8              \\
eastus2            & USA              & Boydton       & 367.8              \\
westus             & USA              & San Francisco & 240.6              \\
northcentralus     & USA              & Chicago       & 568.2              \\
southcentralus     & USA              & San Antonio   & 460.4              \\
northeurope        & Ireland          & Dublin        & 617                \\
westeurope         & Netherlands      & Amsterdam     & 569                \\
japanwest          & Japan            & Osaka-shi     & 516                \\
japaneast          & Japan            & Tokyo         & 516                \\
brazilsouth        & Brazil           & Sao Paulo     & 205                \\
australiaeast      & Australia        & Sydney        & 802                \\
australiasoutheast & Australia        & Melbourne     & 805                \\
southindia         & India            & Pallavaram    & 920                \\
centralindia       & India            & Lohogaon      & 920                \\
westindia          & India            & Mumbai        & 920                \\
canadacentral      & Canada           & Toronto       & 69.3               \\
canadaeast         & Canada           & Quebec        & 20                 \\
uksouth            & United Kingdom   & Midhurst      & 623                \\
ukwest             & United Kingdom   & Wallasey      & 623                \\
westcentralus      & USA              & Mountain View & 297.6              \\
westus2            & USA              & Quincy        & 297.6              \\
koreacentral       & South Korea      & Seoul         & 517                \\
koreasouth         & South Korea      & Busan         & 517                \\
francecentral      & France           & Huriel        & 105                \\
francesouth        & France           & Realmont      & 105                \\
australiacentral   & Australia        & Forrest       & 900                \\
australiacentral2  & Australia        & Forrest       & 900                \\
southafricanorth   & South Africa     & Pretoria      & 1009               \\
southafricawest    & South Africa     & Stellenbosch  & 1009
\end{tabular}
\end{table}

\section*{Appendix B: Hardware Efficiency}\label{appendix}

\begin{table}[h]\label{tab:hardware-efficiency}
\small
\centering
\begin{tabular}{lllllll}
\rowcolor[HTML]{EFEFEF}
\textbf{Name}      & \textbf{Watt (TDP)} & \textbf{TFLOPS32} & \textbf{TFLOPS16} & \textbf{GFLOPS32/W} & \textbf{GFLOPS16/W}           \\ \hline
 & & & & & \\
RTX 2080 Ti        & 250                 & 13.45             & 26.90             & 53.80               & 107.60              \\
RTX 2080           & 215                 & 10.00             & 20.00             & 46.51               & 93.02               \\
GTX 1080 Ti        & 250                 & 11.34             & 0.17              & 45.36               & 0.68                \\
GTX 1080           & 180                 & 8.00              & 0.13              & 44.44               & 0.72                \\
AMD RX480          & 150                 & 5.80              & 5.80              & 38.67               & 38.67               \\
Titan V            & 250                 & 14.90             & 29.80             & 59.60               & 119.20              \\
Tesla V100         & 300                 & 15.00             & 30.00             & 50.00               & 100.00              \\
TPU2               & 250                 & 22.00             & 45.00             & 88.00               & 180.00              \\
TPU3               & 200                 & 45.00             & 90.00             & 225.00              & 450.00              \\
Intel Xeon E5-2699 & 145                 & 0.70              & 0.70              & 4.83                & 4.83               \\
AGX Xavier         & 30                  & 16.00             & 32.00             & 533.33              & 1066.67
\end{tabular}
\end{table}

\end{document}